# A Simple Symmetry as a Guide Towards New Physics Beyond the Standard Model

S. Khalil[1,2,3] and S. Moretti[3,4]

[1]*Center for Theoretical Physics, Zewail City of Science and Technology, Sheikh Zaid, 12588 Giza, Egypt.*

[2]*Department of Mathematics, Faculty of Science, Ain Shams University, Abbassia, 11566, Cairo, Egypt.*

[3]*School of Physics and Astronomy, University of Southampton, Highfield, Southampton SO17 1BJ, UK.*

[4]*Particle Physics Department, Rutherford Appleton Laboratory, Chilton, Didcot, Oxon OX11 0QX, UK.*

*Abstract*
*There exists one experimental result that cannot be explained by the Standard Model (SM), the current theoretical framework for particle physics: non-zero masses for the neutrinos (elementary particles travelling close to light speed, electrically neutral and weakly interacting). The SM assumes that they are massless. Therefore, particle physicists are now exploring new physics beyond the SM. There is strong anticipation that we are about to unravel it, in the form of new matter and/or forces, at the Large Hadron Collider (LHC), presently running at CERN. We discuss a minimal extension of the SM, based on a somewhat larger version of its symmetry structure and particle content, that can naturally explain the existence of neutrino masses while also predicting novel signals accessible at the LHC, including a light Higgs boson, as evidenced by current data.*

The fact that neutrinos have mass is now firm observational evidence for new physics beyond the Standard Model (SM). In the latter, neutrinos are strictly massless due to the absence of right-handed neutrinos and exact global Baryon minus Lepton (B-L) number conservation. This letter considers an extension of the SM, based on the gauge group $SU(3)_C \times SU(2)_L \times U(1)_Y \times U(1)_{B-L}$. Here, we show that it provides a viable and testable solution to the neutrino mass mystery of contemporary particle physics. We also emphasize that it contains a SM-like Higgs boson compatible with current data as well as predicts new particles: heavy neutrinos, that contribute to light neutrino masses, an extra gauge boson associated with the $U(1)_{B-L}$ gauge group, and a new heavier Higgs boson that spontaneously breaks the B-L symmetry at the TeV scale. We argue that all these new states can promptly be probed at the Large Hadron Collider (LHC).

This machine, hosted at the European Organization for Nuclear Research (CERN), with some supplemental help from the Tevatron at the Fermi National Accelerator Laboratory (FNAL), has apparently discovered a Higgs boson [1-3] (which we label H), the only missing, and arguably most important, piece of the current theoretical framework for particle physics, the so called SM. While this discovery represents a milestone in the history of particle physics, it does very little to answer several fundamental questions raised by experimental data. Amongst these, a pressing one emerges from neutrino oscillations, which poses an unresolvable puzzle to the SM, as they clearly provide evidence of non-zero masses for the neutrinos. The SM assumes that its (left-handed) neutrinos are strictly massless due to the absence of (right-handed) neutrinos and an exact (i.e., unbroken) global (i.e., space-time independent) B-L number conservation. (However, at the non-renormalisable level, neutrino masses can be generated within the SM via a dimension 5 operator [4]: in fact, the lepton number is an accidental symmetry in the SM at the renormalisable level, therefore such an operator is not forbidden.)

The need to find answers Beyond the SM (BSM) thus remains clear. In exploring the possibility of some BSM scenario, two key aspects should be borne in mind. Firstly, the very precise experimental confirmations of the SM that have kept accumulating throughout the last four decades make it mandatory for any theory of new physics to exactly reproduce the SM up to the Electro-Weak (EW) energy scale, of O(100 GeV). Secondly, the tremendous success of gauge symmetries in describing Nature, which the SM relies upon, implies a general belief that any new physics should be based on an enlarged gauge group.

Now, two further considerations, which will eventually intertwine, ought to be made. On the one hand, the most attractive dynamics that can naturally account for small yet sizable neutrino masses is known as the seesaw mechanism (in various forms) [5,6]. In this case, three heavy singlet (right-handed) neutrinos are invoked, with super-heavy masses, of order $10^{13}$ Giga electron-Volt (GeV). Although this scenario explains in a rather elegant way why neutrinos are much lighter than the other elementary fermions, it is assumed that it cannot have a direct low energy signature. On the other hand, the aforementioned (B-L) symmetry needs be neither exact nor global. In fact, if both such assumptions are dismissed at once, a whole new dynamics is generated, for the mere purpose of self-consistency of the new model. In practice, if the (B-L) symmetry is locally gauged, then the existence of three SM singlet fermions (the aforementioned right-handed neutrinos) is a quite natural assumption to make in order to cancel the associated triangle anomaly, which is a necessary condition for the self-consistency of the ensuing model [6].

Now, in general, the energy scale of (B-L) symmetry breaking is unknown, ranging from O(TeV) to much higher scales. However, it has been proven in [7] that, in a Supersymmetry (SUSY) framework [8], it is naturally correlated with that of soft SUSY breaking, which is indeed at TeV energies. Therefore, all such new dynamics does not perturb establish physics at the EW scale, yet the predicted new particles, from SUSY or not, will lead to novel signatures at experiments currently probing the TeV energy regime (i.e., the LHC). Further, after (B-L) gauge symmetry breaking has taken place, right-handed neutrinos acquire a mass $M_R$, which can be of O(TeV). Once standard EW Symmetry Breaking (EWSB) has also occurred, at a lower energy, a Dirac neutrino mass $m_D$ is finally generated. Therefore, the mass of the physical light neutrino is given by $m^2_D/M_R$, which can account for the measured experimental results on neutrino oscillation if $m_D \sim 10^{-4}$ GeV [6]. While obviously small, the latter value is clearly not unnatural, as, *e.g.*, the electron mass is $\sim 0.5 \times 10^{-3}$ GeV.

Despite the small mixing between light and heavy neutrinos, new interaction terms between the physical heavy neutrinos (plus the associated leptons) and the weak gauge bosons of the SM (W and Z) are induced. Further, the model also contains an extra gauge boson (hereafter denoted by Z'), corresponding to a now broken (B-L) gauge symmetry, and an extra SM singlet Higgs scalar (denoted by H'), responsible for it, which is heavier than its SM counterpart. So, the phenomenological consequences are numerous. Firstly, the lightest heavy neutrinos can now be copiously (pair) produced via Z' exchange at the LHC [9,10]. Secondly, the Higgs sector of the model (hence the specific light H signal accessible at the LHC, hinting a mass of 125 GeV) is now perturbed by the presence of a heavy state, the H' [11]-[13]. Further, in this class of models, in addition to SM-like decay channels, either or both Higgs bosons can decay in genuine (B-L) final states, like heavy neutrino and/or Z' pairs, with sizable rates. This opens up then the intriguing possibility of all the new states predicted by such a (B-L) model being *simultaneously* detected at the LHC [13] (see also [14]-[16]).

To stay with the Higgs sector, curiously enough, the same LHC data that revealed a Higgs boson also hint at the possibility that this state is not the SM one, because of a clear enhancement of the di-photon production rate, over and above the SM predictions (assuming a Higgs boson). Here, a version of our (B-L) model supplemented by SUSY may have some light to shed. Firstly, the current experimental hint of a SM-like Higgs boson with mass of 125 GeV is uncomfortably very near the absolute upper limit predicted theoretically by the minimal SUSY model (i.e., the one without additional heavy neutrinos, Z' boson and singlet Higgs state plus SUSY counterparts), of 130 GeV or so, so as to seem a very fine-tuned solution. Not so, though, in the SUSY version of the (B-L) model, henceforth the (B-L)SSM. A striking example here is the case of the (B-L)SSM with inverse seesaw [17], whereby, as shown in [18], the one-loop radiative corrections to the lightest SM-like Higgs boson mass, due to the right-handed neutrinos and sneutrinos (their SUSY counterparts) can give an absolute upper limit on it at around 170 GeV. (Needless to say, the possibility that the SM Higgs state had the observed mass would be merely a coincidence, as such a mass is a free parameter). Secondly, the (B-L)SSM can play a crucial role in the explanation of such an enhanced di-photon decay rate, thanks to peculiar contributions to it due to very light staus (the SUSY counterparts of the tau leptons), whose origin is similar to the one in more minimal SUSY models than the (B-L)SSM, yet it occurs under conditions that would more naturally appear at the scale of a Grand Unification Theory (GUT), which physicists are constantly seeking and which they believe being underpinned by SUSY [19].

In short, a symmetry structure deeply rooted in the SM could well be the key to extend the latter into a credible new physics scenario, embedding naturally the neutrino mass patterns measured by experiment and at the same time offering a wealth of new physics signals, all promptly accessible at the LHC, as the dynamics generating the new states occurring in the model can emerge at the TeV scale (and particularly so in its SUSY version), hence well within the reach of the CERN collider, and can easily accommodate a light Higgs boson with its possible anomalies, as evidenced by recent LHC data.

*Acknowledgements* S. Khalil is grateful to The Leverhulme Trust for financial support in the form of a Visiting Professorship. S. Moretti is supported in part through the NExT Institute.